\documentclass[12pt]{article}
\usepackage[english]{babel}
\usepackage{graphicx,color} 
\usepackage{bm}       
\usepackage{amsmath}
\usepackage{amssymb}
\usepackage{caption}
\usepackage[affil-it]{authblk}

\DeclareMathAlphabet{\mathpzc}{OT1}{pzc}{m}{it}
\newcommand{\ket}[1]{ | #1 \rangle}

\begin{document}

\title{The pion-cloud contribution\\ to the electromagnetic nucleon structure}
\author{D. Kupelwieser and W. Schweiger}
\affil{Institute of Physics, University of Graz, A-8010 Graz,
Austria}
\maketitle

\noindent {\bf Abstract}  The present contribution continues and extends foregoing work on the calculation of electroweak form factors of hadrons using the point-form of relativistic quantum mechanics. Here we are particularly interested in studying pionic effects on the electromagnetic structure of the nucleon. To this aim we employ a hybrid constituent-quark model that comprises, in addition to the $3q$ valence component, also a $3q$+$\pi$ non-valence component. With a simple wave function for the $3q$ component we get reasonable results for the nucleon form factors. In accordance with other authors we find that the pionic effect is significant only below $Q^2\lesssim 0.5$~GeV$^2$.

\bigskip

\bigskip

\noindent In a series of papers \cite{Biernat:2009my,Biernat:2010tp,GomezRocha:2012zd,Biernat:2014dea} we have developed and advocated a method for the calculation of the electroweak structure of few-body bound states that is based on the point form of relativistic quantum mechanics. All types of interactions are introduced in a Poincar\'e-invariant way via the Bakamjian-Thomas construction~\cite{Bakamjian:1953kh}. Our strategy is then to determine the invariant $1$-$\gamma$-exchange ($1$-$W$-exchange) amplitude, extract the electroweak current of the bound state, analyze its covariant structure and determine the form factors. The dynamics of the exchanged gauge boson is thereby fully taken into account by means of a coupled-channel formulation.

Here we are interested in the electromagnetic structure of the nucleon as resulting from a hybrid constituent-quark model in which the nucleon is not just a $3$$q$ bound state, but contains, in addition, a $3q$+$\pi$ non-valence component. Transitions between these two components can happen via emission and absorption of the pions by the quarks. In addition, quarks are subject to an {\em instantanous} confining force.
Typical for the point-form, all four components of the momentum operator are interaction dependent, whereas the generators of Lorentz transformations stay free of interactions.  This entails simple rotation and boost properties and angular-momentum addition works like in non-relativistic quantum mechanics. The point-form version of the Bakamjian-Thomas construction allows to separate the overall motion of the system from the internal motion in a neat way:
\begin{equation}\label{eq:massop}
\hat{P}^{\mu}=\hat{\mathcal M}\,  \hat V^{\mu}_{\mathrm{free}}=
\left(\hat{\mathcal M}_{\mathrm{free}}+ \hat{\mathcal
M}_{\mathrm{int}} \right) \hat V^{\mu}_{\mathrm{free}}\, ,
\end{equation}
i.e. the 4-momentum operator factorizes into an interaction-dependent mass operator $\hat{\mathcal{M}}$ and a free 4-velocity operator $ \hat V^{\mu}_{\mathrm{free}}$. Bakamjian-Thomas-type mass operators are most conveniently represented in terms of velocity states $\vert V;  {\bf k}_1, \mu_1; {\bf k}_2, \mu_2; \dots ; {\bf k}_n, \mu_n\rangle$, which specify the system by its overall velocity $V$ ($V_\mu V^\mu=1$), the CM momenta ${\bf k}_i$ of the individual particles and their (canonical) spin projections $\mu_i$~\cite{Biernat:2010tp}.

We now want to calculate the $1$$\gamma$-exchange amplitude for elastic electron scattering off a nucleon that consists of a $3q$ and a $3q$+$\pi$ component. A multichannel formulation that takes not only the dynamics of electron and quarks, but also the dynamics of the photon and the pion fully into account has to comprise all states which can occur during the scattering process (i.e. $|3q, e \rangle$, $|3q, \pi, e \rangle$, $|3q, e, \gamma \rangle$, $|3q, \pi, e, \gamma \rangle$). What one then needs, in principle, are scattering solutions of the mass-eigenvalue equation
\begin{equation}\label{EVequation}
\left(\begin{array}{cccc}
\hat{M}_{3qe}^{\mathrm{conf}} & \hat{K}_\pi & \hat{K}_\gamma & 0
\\
\hat{K}_\pi^\dagger & \hat{M}_{3q \pi e} ^{\mathrm{conf}}& 0 &
\hat{K}_\gamma \\
\hat{K}_\gamma^\dagger & 0 & \hat{M}_{3q e \gamma} ^{\mathrm{conf}}&
\hat{K}_\pi \\
0 & \hat{K}_\gamma^\dagger & \hat{K}_\pi^\dagger &
\hat{M}_{3q \pi e \gamma}^{\mathrm{conf}}
\end{array}\right)
\left(\begin{array}{l}
\ket{\psi_{3q e}} \\ \ket{\psi_{3q \pi e}} \\
\ket{\psi_{3q e \gamma}} \\ \ket{\psi_{3q \pi e \gamma}}
\end{array}\right)
=
\sqrt{s} \left(\begin{array}{l}
\ket{\psi_{3q e}} \\ \ket{\psi_{3q \pi e}} \\
\ket{\psi_{3q e \gamma}} \\ \ket{\psi_{3q \pi e \gamma}}
\end{array}\right)\, ,
\end{equation}
which evolve from an asymptotic electron-nucleon in-state $\ket{e N}$ with invariant mass $\sqrt{s}$. The diagonal entries of this matrix mass operator contain, in addition to the relativistic kinetic energies of the particles in the particular channel, an instantaneous confinement potential between the quarks. The off-diagonal entries are vertex operators which describe the transition between the channels. In the velocity-state representation these vertex operators are directly related to  usual quantum-field theoretical interaction-Lagrangean densities~\cite{Biernat:2010tp}. Since we only deal with pseudoscalar pion-quark coupling in the following, we have neglected the $\pi \gamma q q$-vertex (that would show up for pseudovector pion-quark coupling).

\begin{figure*}[!h]
\centering
\begin{minipage}{0.25\textwidth}
\vspace{-1.5em}
\includegraphics[width=0.9\textwidth]{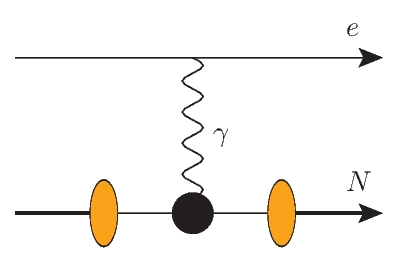}
\end{minipage}
\begin{minipage}{1ex}
\vspace{-1em}
+
\end{minipage}
\begin{minipage}{0.30\textwidth}
\includegraphics[width=0.9\textwidth]{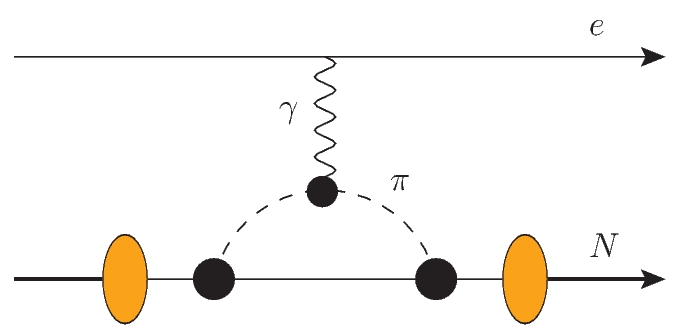}
\end{minipage}
\begin{minipage}{1ex}
\vspace{-1em}
+
\end{minipage}
\begin{minipage}{0.30\textwidth}
\vspace{-1.0em}
\includegraphics[width=0.9\textwidth]{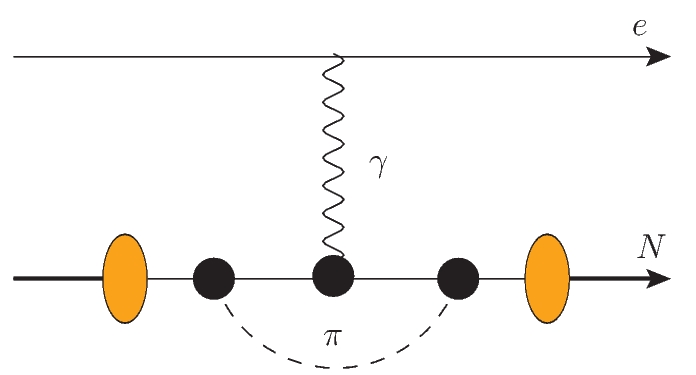}
\end{minipage}
\caption{Diagrams representing the $1$$\gamma$-exchange amplitude for electron scattering off a \lq\lq physical\rq\rq\ nucleon $N$, i.e. a bare nucleon dressed by a pion cloud. The time orderings of the $\gamma$-exchange are subsumed under a covariant photon propagator. Black blobs represent vertex form factors for the coupling of a photon or pion to the bare nucleon $\tilde N$. A vertex form factor is also assumed at the photon-pion vertex. The ovals represent the wave function (i.e. essentially the square root of the probability $P_{\tilde N/N}$) for finding the bare nucleon in the physical nucleon. \label{overallff}}
\end{figure*}

 To proceed, we reduce Eq.~(\ref{EVequation}) to an eigenvalue problem for $\ket{\psi_{3q e}} $ by means of a Feshbach reduction,
\begin{equation}\label{eq:Mphys}
\left[\hat{M}_{3qe}^{\mathrm{conf}} +\hat{K}_\pi(\sqrt{s}-\hat{M}_{3q\pi e}^{\mathrm{conf}} )^{-1} \hat{K}_\pi^\dag + \hat{V}_{1\gamma}^{\mathrm{opt}}(\sqrt{s})\right] \ket{\psi_{3q e}} = \sqrt{s} \, \ket{\psi_{3q \pi e}} \, ,
\end{equation}
where $\hat{V}_{1\gamma}^{\mathrm{opt}}(\sqrt{s})$ is the 1$\gamma$-exchange optical potential. The invariant 1$\gamma$-exchange electron-nucleon scattering amplitude is now obtained by sandwiching $\hat{V}_{1\gamma}^{\mathrm{opt}}(\sqrt{s})$ between (the appropriately normalized valence component of the) physical electron-nucleon states  $\ket{eN}$, i.e. eigenstates of $[  \hat{M}_{3qe}^{\mathrm{conf}} +\hat{K}(\sqrt{s}-\hat{M}_{3q\pi e}^{\mathrm{conf}} )^{-1} \hat{K}^\dag ]$. The crucial point is now to observe that, due to instantaneous confinement, propagating intermediate states do not contain free quarks, they rather contain either physical nucleons  $N$ or bare baryons $\tilde{B}$, the latter being eigenstates of the pure confinement problem. As a consequence one can reformulate the scattering amplitude in terms of pure hadronic degrees of freedom with the quark substructure being hidden in vertex form factors. This is graphically represented in Fig.~\ref{overallff}. The analytical expressions for the bare nucleon form factors follow from this reformulation.\footnote{For details of their extraction and a discussion of the problems connected with wrong cluster properties associated with the Bakamjian-Thomas construction we refer to Refs.~\cite{Biernat:2009my,Biernat:2010tp,Biernat:2014dea}.}

If nucleonic excitations are neglected in the pion loop, we just need the electromagnetic $\gamma\tilde{N}\tilde{N}$ and strong $\pi\tilde{N}\tilde{N}$ vertex form factors (for the bare nucleon $\tilde N$) as well as the electromagnetic pion form factor.
The electromagnetic pion form factor can be taken from Ref.~\cite{Biernat:2009my}, where it has been calculated within the same approach as here using a harmonic-oscillator model for the $u\bar d$ bound-state wave function of the $\pi^+$. What enters the analytical expressions for the form factors of the bare nucleon is its $3$$q$ bound-state wave function. Instead of solving the bound-state problem for a particular confinement potential, we rather use a simple model for this wave function, i.e. $\Phi_{\tilde N}\,\left({\tilde {\bf k}}_i\right)={\mathcal{N}}{\left[(\sum \tilde{\omega}_i)^2 + \beta^2\right]^{-\gamma}}\, ,
$ with ${\tilde {\bf k}}_i$ and $\tilde{\omega}_i$ denoting the quark momenta and energies in the rest frame of the nucleon. The same wave function has been used in a corresponding front-form calculation~\cite{Pasquini:2007iz}, from which we also take the values of the parameters $\beta$, $\gamma$ for later comparison. The normalization $\mathcal{N}$ has to be fixed such that the whole nucleon wave function, including the $3q$+$\pi$ component, is normalized to one. Unlike the authors of Ref.~\cite{Pasquini:2007iz}, who took a phenomenological $\pi\tilde{N}\tilde{N}$ vertex form factor, we have calculated both, the electromagnetic form factors of the bare nucleon as well as the strong $\pi\tilde{N}\tilde{N}$ vertex form factor with the same microscopic input, namely the $3q$ bound-state wave function $\Phi_{\tilde{N}}$.

\begin{figure}
\centering
\includegraphics[width=0.48\textwidth]{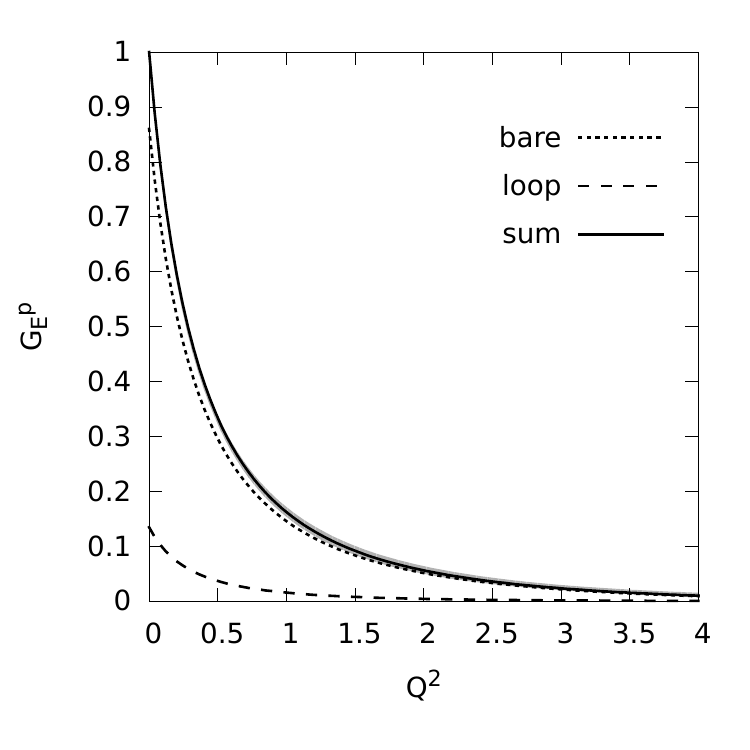}
\includegraphics[width=0.48\textwidth]{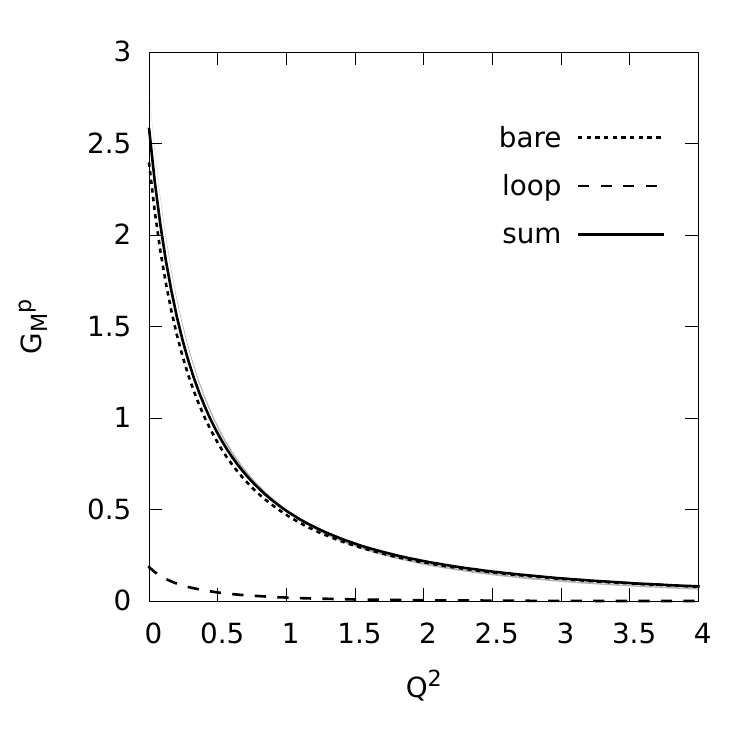}\vspace{-0.7cm}
\caption{The proton electric (left) and magnetic (right) form factors as predicted by our model (solid line). The $3q$ valence contribution is indicated by the dotted line, the contribution due to the $3q$+$\pi$ non-valence component by the dashed line. The shaded area (which is hardly visible) is a parameterization of the experimental data (including uncertainties)~\cite{Puckett:2010kn}.} \label{fig:protonffs}
\end{figure}

With the model sketched above we achieve good agreement with the experimental data for proton electric and magnetic form factors (see Fig.~\ref{fig:protonffs}). Our neutron magnetic form factor is also in reasonable agreement with the corresponding experimental data, the reproduction of the neutron electric form factor seems to be less satisfactory. But here one has to notice that it is a rather small quantity and the error bars on the experimental data points are, in general, large. The  size of the pionic contribution to all the nucleon form factors is comparable with the one found in Ref.~\cite{Pasquini:2007iz}. A significant effect of the $3q$+$\pi$ component on the form factors is only observed for  momentum transfers $Q^2\lesssim 0.5$~GeV$^2$, where it leads to a welcome modification of the $Q^2$-dependence.

\medskip
Improvements of the model can be made in several directions:
\begin{itemize}
\vspace{-0.2cm}
\item[i)] Take a more sophisticated $3q$ wave function of the (bare) nucleon, containing, e.g., a mixed $SU(6)$ spin-flavor-symmetry component like in Ref.~\cite{Pasquini:2007iz}.
\vspace{-0.2cm}
\item[ii)] Replace the pseudoscalar by the pseudovector $\pi \tilde{N}\tilde{N}$ coupling, which guarantees correct properties in the chiral limit.
\vspace{-0.2cm}
\item[iii)]  Acount for other baryons, different from the nucleon, within the loop, the lightest and most important of them being the $\Delta$.
\end{itemize}
The ultimate goal should, of course, be a consistent description of the baryon spectrum and the structure of the baryons. This means that one should not start with a model for the nucleon wave function, but rather with a $3q$+$3q\pi$ hybrid model and fit the parameters of the confinement potential and the $\pi q q$ coupling strength to the baryon mass spectrum. This would give us the masses and wave functions of the (bare) baryons which are required as an input for the calculation of the strong and electromagnetic form factors of the baryons.

\quad

\noindent{\bf Acknowledgement:}
D. Kupelwieser acknowledges the support of the \lq\lq Fonds zur
F\"orderung der wissenschaftlichen Forschung in \"Osterreich\rq\rq\ (FWF
DK W1203-N16).

%
%
%

\end{document}